\documentclass[11pt]{article}

\usepackage{mymacros}

\usepackage{cite}
\title{\textbf{All near-horizon symmetries of the Schwarzschild black hole in linearised gravity}}
\author{Ankit Aggarwal$^{\star , \ddag}$ and Nava Gaddam$^{\dagger}$}
\affiliation[]{$^{\star}$Institute for Theoretical Physics Amsterdam and Delta Institute for Theoretical Physics, \\ University of Amsterdam, Science Park 904, 1098 XH Amsterdam, The Netherlands.}
\affiliation[]{$^{\ddag}$Physique Math\'{e}matique des Interactions Fondamentales, Universit\'{e} Libre de Bruxelles and \\ International Solvay Institutes, Campus Plaine - CP 231, 1050 Bruxelles, Belgium.}
\affiliation[]{$^{\dagger}$ International Centre for Theoretical Sciences, Tata Institute of Fundamental Research, \\ Shivakote, Bengaluru 560089, India.} 
\emailAdd{ankit.aggarwal@ulb.be}
\emailAdd{n.gaddam@icts.res.in}
\abstract{Asymptotic symmetries are known to constrain the infrared behaviour of scattering processes in asymptotically flat spacetimes. By the same token, one expects symmetries of the black hole horizon to constrain near-horizon gravitational scattering. In this paper, we take a step towards establishing this connection. We find all near-horizon symmetries that can be potentially relevant to gravitational scattering near the horizon of the Schwarzschild black hole. We study large diffeomorphisms of linearised perturbations of the Schwarzschild black hole in a partial wave basis and in a gauge that allows for gravitational radiation crossing the event horizon. This setup is ideally suited for studying processes involving near-horizon gravitons like scattering and black hole evaporation. We find the most general near-horizon symmetries that are consistent with finite perturbations at the horizon. Since we do not impose any further boundary conditions, these symmetries represent the biggest set of symmetries in this setting. We find the associated covariant charges to be finite and non-zero showing that these symmetries are physical. Interestingly, for a large black hole, the dominant symmetries are just two copies of $ u(1)$. }

\date{\today}

\begin{document}
	
	\maketitle
	
	\section{Introduction}
	Classical analyses of perturbations of the Schwarzschild black hole have a rich history. Seeking an understanding of the classical stability of the background, influential work on the subject dates back to several decades ago \cite{Regge:1957td, Vishveshwara:1970cc, Zerilli:1970wzz}. These lead to further important developments due to Chandrasekhar \cite{Chandrasekhar:1985kt}. In addition to various applications \cite{Frolov:1998wf, Gleiser:1998rw, Sasaki:2003xr, Martel:2003jj}, an effort to cast the perturbations in gauge invariant form has culminated in a rather practical and covariant avatar \cite{Moncrief:1974am, Gerlach:1979rw, Sarbach:2001qq, Cunningham:1978zfa, Jhingan:2002kb, Martel:2005ir} which also allowed for a study of classical gravitational radiation both crossing the horizon and reaching infinity.
	
	Owing to Hawking's seminal work \cite{Hawking:1975vcx, Hawking:1976ra}, quantum aspects of black holes have an equally captivating and yet, as is widely acknowledged, an incomplete story. While the problem of information loss has been widely debated, quantum aspects of the most general perturbations of the Schwarzschild black hole have received considerably little attention. Only recently has a formalism to study scattering processes in the near-horizon region been developed \cite{Gaddam:2020rxb, Gaddam:2020mwe, Betzios:2020xuj, Gaddam:2021zka, Gaddam:2022pnb}.\footnote{For previous work that led to these developments, see \cite{tHooft:1996rdg, Hooft:2015jea, Hooft:2016itl, Betzios:2016yaq}.}  These processes naturally include metric perturbations resulting in a new regime of quantum gravity where elastic $2-2$ amplitudes eikonalise. This regime is where centre of mass energies, $ E$, of scattering processes satisfy $E M \gg M^{2}_{Pl}$ \cite{Gaddam:2020rxb, Gaddam:2020mwe}, where $M$ is the mass of the black hole. The formalism also allows for calculations of S-matrix elements away from the black hole eikonal regime. Inelastic processes relevant for black hole evolution can also be explicitly calculated \cite{Gaddam:2021zka}. A new soft limit is also expected to emerge in the near-horizon limit where graviton momenta scale inversely with the Schwarzschild radius in the large black hole limit \cite{us}. These developments are inherently based on the covariant avatar of the black hole perturbations alluded to earlier.
	
	Black holes notwithstanding, it has become increasingly apparent that the infrared structure of gravity is far richer than previously thought even perturbatively about flat space. Gravitational radiation reaching null infinity is known to be intricately tied to infinite dimensional symmetries arising from large gauge transformations that survive at future and past null infinities \cite{Sachs1962, Sachs1962a, Bondi:1962px}. In turn, a diagonal subgroup of these future and past Bondi-Metzner-Sachs (BMS) groups has been argued to be a symmetry of the asymptotically flat quantum gravity S-matrix \cite{Strominger:2013jfa}. The Ward identities associated with these symmetries have been shown to be the same as Weinberg's soft graviton theorem in flat space \cite{Weinberg:1965nx, He:2014laa, Strominger:2017zoo}. 
	
	It is then natural to ask if the radiation crossing the horizon has an analogously rich relationship with asymptotic symmetries of the horizon and if these are tied to  the near-horizon scattering processes. Infinite dimensional symmetries have in fact been shown to arise near the horizon of non-extremal black holes \cite{Donnay:2015abr, Hawking:2016msc, Donnay:2016ejv, Hawking:2016sgy,  Haco:2018ske,Aggarwal:2019iay, Chandrasekaran:2018aop, Grumiller:2019fmp, Adami:2020amw, Adami:2021nnf, Chandrasekaran:2021vyu,Freidel:2021cjp,Ciambelli:2021vnn,Sheikh-Jabbari:2022mqi, Chandrasekaran:2023vzb, Odak:2023pga }. Their relevance to the black hole information problem has also been debated \cite{Mirbabayi:2016axw, Strominger:2017aeh}. These symmetries can be thought as being emergent near the horizon, once the collapse process of a large black hole has settled. They are valid for as long as the black hole is semi-classically large ($M\gg M_{Pl}$). It is not known if these symmetries in the literature are the complete set of all symmetries of the horizon. Furthermore, any potential relationship between these symmetries and scattering processes near the horizon is also a glaringly open problem.
	
	\paragraph{Results: } In this article, we take an important step towards addressing the questions raised in the previous paragraph. Within the covariant approach to black hole perturbations, we derive the set of all near- horizon symmetries assuming only that the perturbations remain finite on the horizon.  Remarkably, with no further boundary conditions on the perturbations than their  finiteness  at the horizon, we find the corresponding asymptotic charges to be finite. The asymptotic Killing vectors thus found, form a closed abelian algebra at the linearised level but there are non-linear obstructions to closure. These must be taken care of by modifying the Lie bracket, \textit{a la} Barnich-Troessaert \cite{Barnich:2011mi}, in a non-linearised analysis which is beyond the scope of this work. However, if one demands the non-linear closure without modifying the Lie bracket, the maximal sub-algebra turns out to arise from more stringent boundary conditions that set certain non-radiative data to zero. The resulting sub-algebra contains $\text{Diff}\left(S^{2}\right)$ in semi-direct sum with two supertranslations. In fact, we find the Killing vectors to all orders in the near-horizon expansion parameter.  In the large black hole limit, the dominant symmetries form two copies of $ u(1) $, one of which is not contained in the aforementioned maximal sub-algebra. 
	
	\paragraph{Organisation of this paper: } In \secref{sec:setup}, we set up the perturbations of the Schwarzschild black hole in a partial wave basis and explain our choice of gauge that allows for gravitational radiation crossing the horizon. In \secref{sec:symmetriesandcovcharges}, we derive the  residual diffeomorphisms that preserve the said gauge choice. We also impose regular boundary conditions for the perturbations on the horizon to find the near-horizon symmetries, and work out the associated charges using covariant phase space formalism. We also discuss the physical interpretation of the charges and show that two of them represent black hole entropy and angular momentum. In \secref{sec:algebra}, we turn to the algebra of these near-horizon Killing vectors and identify the sub-algebras that close even at the non-linear level without any modification to the Lie bracket. We also study the symmetries relevant for large black holes.  Finally, we consider the extension of our near-horizon symmetries to all orders in the distance away from the horizon.  We conclude with a brief summary of results and some open questions in \secref{sec:conclusions}.

	\section{Black hole perturbations in partial waves}\label{sec:setup}
	We are interested in metric perturbations, $ h_{\mu \nu} $, defined via ${g}_{\mu\nu} ~ = ~ \bar g_{\mu\nu} + \kappa h_{\mu\nu}$ with $\kappa^2 = 8\pi G$. The background is the Schwarzschild black hole denoted by $\bar g_{\mu\nu}$ that is specified by
	\begin{equation}\label{eqn:background}
		\mathrm{d}s^{2} ~ = ~ A\left(u,v\right) \mathrm{d} u \mathrm{d} v + r\left(u,v\right)^{2} \mathrm{d}\Omega^{2} \quad \text{where} \quad A\left(u,v\right) = \dfrac{R}{r} \exp\left(1 - \dfrac{r}{R}\right) \, ,
	\end{equation}
	where $ R $ is the Schwarzschild radius and $r\left(u,v\right)$ is implicitly determined from the relation:
	\begin{equation}\label{eqn:rofuv}
		uv ~ = ~ 2 R^{2} \left(1 - \dfrac{r}{R}\right) \exp\left(\dfrac{r}{R} - 1\right) \, .
	\end{equation}
	The horizon is located at $ r=R $ implying that either $ u=0 $ or $ v=0 $. We call $ u=0 $ and  $ v=0 $ as the future and past horizon, respectively. While we restrict our attention to the past horizon (wherefrom outgoing radiation emanates) in this article, analogous considerations directly apply to the future horizon.
	
	Perturbations of the Schwarzschild black hole are most naturally studied in the partial wave decomposition\footnote{In this paper, following \cite{Gaddam:2020rxb, Gaddam:2020mwe}, we use the real representation of the spherical harmonics which satisfy the same orthogonality relations as the complex ones.} of Regge and Wheeler \cite{Regge:1957td}. In this basis, metric perturbations are split into the so-called even and odd modes, respectively, as follows:
	\begin{subequations}\label{eqn:gravitonModes}
		\begin{align}
			h^+_{\mu\nu} 
			~&= ~ \sum_{\ell, m} ~ \begin{pmatrix}
				H_{ab} & - h^{+}_{a} D_{A} \\
				- h^{+}_{a} D_{A} & ~~~~ r^{2}  \left(K+\frac{\ell(\ell+1)}{2} G\right) \gamma_{AB} + r^{2} G D_{A} D_{B}
			\end{pmatrix} Y_{\ell m} \, , \label{eqn:evenGraviton} \\
			h^-_{\mu\nu} 
				~&= ~ \sum_{\ell, m}~ \begin{pmatrix}
					0 & - h^{-}_{a} {\epsilon_{A}}^{B} D_{B} \\
					- h^{-}_{a} {\epsilon_{A}}^{B} D_{B} & ~~~~ - h_{\Omega} {\epsilon_{(A}}^{C} D_{B)} D_{C}
				\end{pmatrix}  Y_{\ell m} \, . \label{eqn:oddGraviton}
			\end{align}
		\end{subequations}
		All metric components in this decomposition naturally carry partial wave indices which we have suppressed to avoid a clutter of notation. We will continue to leave the partial wave indices implicit except when we choose to remind ourselves of their presence. The longitudinal Kruskal coordinates $(u, \, v)$ are labelled by lowercase Latin indices, while the corresponding uppercase ones stand for the angular coordinates. Moreover, $D_{A}$ stands for the covariant derivative on the unit two-sphere parametrised by the round metric $\gamma_{AB}$. Finally, ${\epsilon_{A}}^{B}$ is the completely antisymmetric tensor on the sphere with ${\epsilon_{\theta}}^{\phi} = \sin\theta$.

		\subsection{Radiation gauge}
		Consider a generic vector field 
		decomposed into vector spherical harmonics as
		\begin{align}\label{eqn:genericDiffPWs}
			\bar{\chi}_{a} ~ = ~ \kappa\sum_{\ell m} \chi^{\ell m}_{a} Y_{\ell m} \quad \text{and} \quad \bar{\chi}_{A}
			~ = ~ \kappa\sum_{\ell m} \left( \chi^{+}_{\ell m} \partial_{A} + \chi^{-}_{\ell m} {\epsilon_{A}}^{B} \partial_{B} \right) Y_{\ell m} \, .
		\end{align}
		It is evident that three of the components of the vector field are even modes while $\chi^{-}_{\ell m}$ is an odd mode. As discussed in \cite{Gaddam:2020rxb, Gaddam:2020mwe, Kallosh:2021ors, Kallosh:2021uxa, Gaddam:2021zka}, a convenient choice of gauge (the ``Regge-Wheeler gauge'') that leaves no residual gauge symmetry is one where the vector fields are chosen such that $G = 0$, $h_{\Omega} = 0$, $h^{+}_{a} = 0$. This ensures that the even perturbations  are block diagonal whereas the odd perturbations are entirely off-diagonal. However, this gauge does not allow for any radiative data. 
		
		We will use a gauge that allows for gravitational radiation crossing the past horizon, called the ``radiation gauge'', first proposed in \cite{Martel:2005ir}:
		\begin{equation}\label{eq:radiationgauge}
			H_{uv} ~ = ~ 0 \, , \quad H_{uu} ~ = ~ 0 \, , \quad \text{and} \quad h_u^\pm ~ = ~ 0 \, .
		\end{equation}
		A similar condition can be imposed to study radiation on the future horizon and analogous results can be obtained. Near the past (future)  horizon, $ v=0~ (u=0) $, it was argued in \cite{Martel:2005ir} that the free radiative data can be chosen to be the leading component of $ G $ when expanded in a small $ v$ series ($u$ series). This component is at $ \mathcal O (v^{0}) $. Moreover, as pointed out in \cite{Martel:2005ir, Gaddam:2020rxb, Gaddam:2020mwe} and further exploited in \cite{Kallosh:2021ors, Kallosh:2021uxa, Gaddam:2021zka}, it may be worth noting that there are no propagating degrees of freedom in the monopole ($\ell=0$) and dipole ($\ell=1$) sectors. Only the multipole modes with $\ell > 1$  are physical and propagating.

		\section{Symmetries and covariant charges}\label{sec:symmetriesandcovcharges}
		In this section, we find the residual symmetries of the radiation gauge, demand that the perturbations be regular at the horizon as boundary conditions, and compute the near-horizon charge using the covariant phase space method.

		\subsection{Residual gauge symmetry}\label{sec:ResGaugeSymm}
		With the parameterisation \eqref{eqn:genericDiffPWs} of a generic diffeomorphism decomposed in partial waves, the metric components in \eqref{eqn:gravitonModes} transform as
		\begin{subequations}\label{eqn:gravitonVariations}
			\begin{align}
				\delta H^{\ell m}_{ab} ~ &= ~ \tilde{\nabla}_a \chi^{\ell m}_b + \tilde{\nabla}_b \chi^{\ell m}_a ~ , \label{eqn:HabVar} \\
				\delta h_{a, \ell m}^{+} ~ &= ~ \chi^{\ell m}_a + \tilde{\nabla}_a \chi^+_{\ell m} - \dfrac{2}{r} \left(\partial_a r\right) \,\chi^+_{\ell m} ~ , \label{eqn:hplusaVar} \\
				\delta K_{\ell m} ~ &= ~ \dfrac{2}{r} g^{a b} \left(\partial_{a} r\right) \chi^{\ell m}_b - \dfrac{\ell \left(\ell + 1\right) }{r^2} \chi^+_{\ell m} \, , \label{eqn:KVar} \\
				\delta G_{\ell m} ~ &= ~ \dfrac{2}{r^2} \chi^+_{\ell m} ~ , \label{eqn:GVar} \\
				\delta h_{a, \ell m}^{-} ~ &= ~ \nabla_a \chi^-_{\ell m} - \dfrac{2}{r} \left(\partial_a r\right) \, \chi^-_{\ell m} ~ , \label{eqn:hminaVar} \\
				\delta h^{\ell m}_{\Omega} ~ &= ~  2 \chi^-_{\ell m} ~ . \label{eqn:hOmegaVar}
			\end{align}
		\end{subequations}
		Here all the covariant derivatives are with respect to  the background metric and $\tilde{\nabla}$ refers to the covariant derivative restricted to the longitudinal coordinates. We are interested in finding those gauge transformations that keep us in the radiation gauge \eqref{eq:radiationgauge}. The radiation gauge imposes constraints on the diffeomorphisms in \eqref{eqn:genericDiffPWs}. Demanding $H_{uu} ^{\ell m}= 0$ leads to
		\begin{align}
			0 ~ = ~ \tilde{\nabla}_{u} \chi^{\ell m}_{u}
			= ~ \partial_{u} \chi^{\ell m}_{u} - \dfrac{\partial_{u} A\left(r\right)}{A\left(r\right)} \chi^{\ell m}_{u} \, .
		\end{align}
		Therefore, we find that the most general solution for $\chi^{\ell m}_{u}$ in terms of an arbitrary integration constant $f_1^{\ell m}\left(v\right)$ is of the form
		\begin{equation}
			\chi^{\ell m}_{u} ~ = ~ A\left(r\right) f_{1}^{\ell m}\left(v\right) \, .
		\end{equation}
		Next,  we have that $h^{+}_{a, \ell m} = 0$ implies
		\begin{align}
			0 ~ &= ~ \left(\partial_{u} - \dfrac{2}{r} \partial_{u} r\right) \chi^{+}_{\ell m} + \chi^{\ell m}_{u} \, .
		\end{align}
		resulting in the solution
		\begin{align}
			\chi^{+}_{\ell m} ~ = ~ r^{2} \left(f_{2}^{\ell m}\left(v\right) - f_{1}^{\ell m}\left(v\right) \int \mathrm{d} u \, \dfrac{A\left(r\right)}{r^{2}}\right) \, ,
		\end{align}
		where we introduced a new integration constant $f_{2}\left(v\right)$. 
		The last gauge condition, $H_{u v}^{\ell m} = 0$, implies that 
		\begin{align}
			0 
			~ = ~ \partial_{u} \chi^{\ell m}_{v} + \partial_{v} \chi^{\ell m}_{u} \, ,
		\end{align}
		where 
		we used that, in the background \eqref{eqn:background}, $g^{ab} \Gamma^{c}_{a b} = 0$. Therefore, we find the following solution\footnote{Alternatively, the solution to this equation may also be written in terms of an auxillary function $q^{\ell m}\left(u,v\right)$ such that $\chi^{\ell m}_{a} = g^{b c} \epsilon_{a c} \partial_{b} q^{\ell m}$ which implies $\chi^{\ell m}_{v} = - \frac{1}{A\left(r\right)} \partial_{v} q^{\ell m}$.} in terms of yet another integration constant $f_{3}\left(v\right)$:
		\begin{align}
			\chi^{\ell m}_{v} ~ &= ~ f_{3}^{\ell m}\left(v\right) - \partial_{v} f_{1}^{\ell m}\left(v\right) \int \mathrm{d}u \, A\left(r\right) - f_{1}^{\ell m}\left(v\right) \int \mathrm{d}u \, \partial_{v} A\left(r\right) \, .
		\end{align}
		Finally, requiring $h^{-}_{u, \ell m} = 0$ yields the equation
		\begin{align}
			0 ~ &= ~ \partial_{u} \left(\dfrac{\chi^{-}_{\ell m}}{r^{2}}\right) \quad \text{which has the solution} \quad \chi^{-}_{\ell m} ~ = ~ r^{2} f^{\ell m}_{4}\left(v\right) \, .
		\end{align}
		Collecting these components, we find the residual Killing vectors of the radiation gauge to be
		\begin{subequations}\label{eqn:KV}
			\begin{align}
				\chi^{\ell m}_{u} ~ &= ~ A\left(r\right) f_{1}\left(v\right) \, ,\quad \chi^{\ell m}_{v} ~ &&= ~ f_{3}\left(v\right) - \partial_{v} f_{1}\left(v\right) \int \mathrm{d}u \, A\left(r\right) - f_{1}\left(v\right) \int \mathrm{d}u \, \partial_{v} A\left(r\right) \label{eqn:KVuv} \\
				\chi^{-}_{\ell m} ~ &= ~ r^{2} f_{4}\left(v\right) \, , \qquad \chi^{+}_{\ell m} ~ &&= ~ r^{2} \left(f_{2}\left(v\right) - f_{1}\left(v\right) \int \mathrm{d} u \, \dfrac{A\left(r\right)}{r^{2}}\right) \, . \label{eqn:KVpm}
			\end{align}
		\end{subequations}
		The derivation of these equations did not require any near-horizon expansion and they are valid to all orders in $ v $.

		\subsection{Boundary conditions and equations of motion near the horizon}\label{sec:eom}
		The only boundary conditions we demand are that the gravitational perturbations do not diverge at the horizon. As we will see, this rather mild constraint leads to finite near-horizon charges.  The requirement that the perturbations remain finite allows us to expand all the six remaining off-shell fields $\Phi \in \left\{ G, \, K , \, h^{\pm}_{v} , \, H_{vv} , \, h_{\Omega} \right\}$ in a Madhavan-Taylor expansion around  $ v=0 $:
		\begin{equation}\label{eq:NHexpansion}
			\Phi ~ = ~ \sum_{n=0}^{\infty} \Phi^{(n)}v^n ~ .
		\end{equation}
		It may be checked that the $uu$-component of the leading order linearised vacuum equations of motion imply that $ K^{(0)} $ is non-radiative free data determined by two arbitrary constants \footnote{In \cite{Martel:2005ir}, $K^{(0)}$ was set to 0 which turns out to be very constraining for the near-horizon symmetries.} $c_{K^{(0)}}$ and $d_{K^{(0)}}$:
		\begin{equation}\label{eqn:K0eom}
			\partial_u^2K^{(0)} ~ = ~ 0 \quad \text{which implies that} \quad K^{(0)} ~ = ~ c_{K^{(0)}} u + d_{K^{(0)}} \, .
		\end{equation}
		All the other even perturbations can be expressed in terms of $ G^{(0)} $ together with other non-radiative free data
		\begin{equation}\label{eq:eom}
			\begin{aligned}
				\partial_u G^{(1)} ~ &= ~ \dfrac{1}{2 R^2} \left(u \partial_u G^{(0)} + 2 h_v^{(0)} \right)  \\
				\partial_u K^{(1)} ~ &= ~ -\dfrac{1}{2 R^2} \left( \ell \left(\ell + 1\right) \partial_u h_v^{(0)} - u \partial_u K^{(0)} + R^2 \partial_u^2 H_{vv}^{(0)}\right)  \\
				\partial_u^2 h_v^{+(0)} ~ &= ~ -\dfrac{1}{2} \left(\ell - 1\right) \left(\ell + 2\right) \partial_u G^{(0)} - \partial_u K^{(0)}  \\
				\partial_u^2 H_{vv}^{(0)} ~ &= ~ \dfrac{1}{2 R^2}\left ( \left(\ell - 1\right) \ell \left(\ell + 1\right) \left(\ell + 2\right) G^{(0)} + 2 \left( \left(\ell + 2\right) \left(\ell - 1\right) \right) K^{(0)} - 2 u \partial_u K^{(0)} \right) \, ,
			\end{aligned}
		\end{equation}
		The first of these equations arises from the $\theta\phi$ component of the equations of motion, the second from the $\phi\phi$ component, and the third from the $u\phi$ component. Therefore, it is evident that near the  past (future)  horizon, $ v=0 ~ (u=0) $, the free radiative data for the even sector may be chosen to be the leading $\mathcal{O}\left(1\right)$ component of $ G $ when expanded in a small $ v$ series ($u$ series). Of course, the leading components of $h^{+}_{v}$ or $H_{vv}$ are equally valid choices. The above equations can be solved to yield
		\begin{align}\label{eq:eomsol}
			K^{(1)} ~ &= ~ \dfrac{1}{2 R^2} \left[2 c_{K^{(0)}} u^2 - \left\{\ell \left(\ell + 1\right) c_{h} + \left(\ell + 2\right) \left(\ell - 1\right) d_{K^{(0)}} \right\}u \right] + d_{K^{(1)}} ~ , \nonumber \\
			h_v^{+(0)} ~ &= ~ -\dfrac{1}{2} \left(\ell - 1\right) \left(\ell + 2\right) \left(\int \dd u G^{(0)} \right) - \dfrac{u^2}{2} c_{ K^{(0)}} + c_{h} u + d_{h} ~ , \nonumber \\
			\partial_u H_{vv}^{(0)} ~ &= ~ \dfrac{1}{2 R^2} \big[ \left(\ell - 1\right) \ell \left(\ell + 1\right) \left(\ell + 2\right) \left(\int \dd u ~ G^{(0)} \right) + \left(\ell^2 + \ell - 3\right) c_{K^{(0)}} {u^2} \nonumber \\
			&\qquad + ~  2 \left(\ell + 2\right) \left(\ell - 1\right) d_{K^{(0)}} u + d_{H} \big] ~ .
		\end{align}
		Here we introduced additional constants of integration: $c_{h}$, $d_{K^{(1)}}$, $d_{h}$, and $d_{H}$. In the odd sector, on the other hand, the leading order linearised vacuum equations of motion imply 
		\begin{equation}\label{eq:eomodd}
			\partial_u h_{\Omega}^{(0)} ~ = ~ - \dfrac{2 R^2}{\left(\ell + 2\right) \left(\ell - 1\right)} \partial_u^2 h_v^{-(0)} ~ .
		\end{equation}
		All the subleading fields can be expressed in terms of $ h_{v}^{-(0)} $ or $ h_{\Omega}^{(0)} $ together with some non-radiative data. Thus, near the past (future) horizon, $ v=0 ~ (u=0) $, the free radiative data for the odd sector may be chosen to be the leading component of $ h_{\Omega} $ 
		when expanded in a small $ v$ series.

		\subsection{Asymptotic Killing vectors}\label{sec:AKV}
		The near-horizon expansions \eqref{eq:NHexpansion} together with \eqref{eqn:gravitonVariations} lead to the following expansions for the Killing vector near the past horizon 
		\begin{equation}
			\chi^{\ell m}_a ~ = ~ \sum_{n=0}^{\infty} \chi_a^{(n)} v^n ~, \quad \chi^+_{\ell m} ~ = ~ \sum_{n=0}^{\infty} \chi^{+(n)} v^n ~ , \quad  \chi^-_{\ell m} ~ = ~ \sum_{n=0}^{\infty}\chi^{-(n)} v^n ~ ,
		\end{equation}
		where the leading order components can be found to be\footnote{We introduced additional factors of $R$ in some coefficients to ensure that all constants are dimensionless. These may be fixed by noting the dimensions of the metric perturbations in \eqref{eqn:gravitonModes} and deducing the dimensions of the Killing vectors from \eqref{eqn:gravitonVariations}. Based on this analysis, it can be checked that $\left[\chi_{a}\right] = L^{0}$, $\left[\chi^{\pm}\right] = L^{1}$.}
		\begin{subequations}\label{eqn:AKVleadingOrder}
			\begin{align}
				\chi_v^{(0)} ~ &= ~ - \dfrac{\alpha_{1} u^2}{2 R^2} + \dfrac{\alpha_2 u}{R} + \alpha_3 \, , \\
				\chi_u^{(0)} ~ &= ~ \alpha_{1} 
				\, , \\
				\chi^{+(0)} ~ &= ~ - \alpha_{1} u + R \beta_{1} ~ , \\
				\chi^{-(0)} ~ &= ~ R \gamma_1 
			\end{align}
		\end{subequations}
		where the partial-wave dependent constants $\alpha_{i}$, $\beta_{i}$, and $ \gamma_i $ are determined in terms of the integration constants in \eqref{eqn:KV}. Writing all integration constants appearing in \eqref{eqn:KV} also in a power series as $f_{i}(v) = \sum_{n=0}^{\infty} f^{(n)}_{i} v^{n}$, we have that $\alpha_{1} = f^{(0)}_{1}$, $\alpha_{2} = - R f^{(1)}_{1}$, $\alpha_{3} = f_{3}^{(0)}$, $\beta_{1} = R f^{(0)}_{2}$, and $\gamma_1 = R f^{(0)}_{4}$. These solutions follow from \eqref{eqn:KV} where we used that, in the Schwarzschild background, $\partial_{a} A \left(r\right) = \partial_{a} r \partial_{r} A\left(r\right)$ and that 
		\begin{equation}
			r\left(u,v\right) = R - \frac{uv}{2R} -\frac{u^2v^2}{4R^3} +\mathcal{O}\left( v^{3}\right)~, \quad A(u,v)=1+\frac{uv}{R^2}+\frac{9 u^2 v^2}{8 R^2}+\mathcal{O}\left( v^{3}\right)~
		\end{equation}
		near the horizon. The subleading terms can similarly be determined to be 
		\begin{subequations}\label{eqn:AKVsubleadingOrder}
			\begin{align}\label{key} 
				\chi_v^{(1)}(u) ~ &= ~ - \dfrac{3 \alpha_{1}}{4 R^4} u^3 + \dfrac{\alpha_2}{R^3} u^2 + \dfrac{\alpha_4 u}{R^{2}} + \dfrac{\alpha_5}{R} \, , \\ 
				\chi_u^{(1)}(u) ~ &= ~ \dfrac{\alpha_{1} u}{R^2} - \dfrac{\alpha_2}{R} ~ , \\
				\chi^{+(1)}(u) ~ &= ~ \left(\dfrac{\alpha_{2}}{R} - \dfrac{\beta_{1} }{R}\right) u + \beta_2 ~ , \\
				\chi^{-(1)} ~ &= ~ \gamma_{2} - \frac{\gamma_1 u }{R}~ .
			\end{align}
		\end{subequations}
		The new coefficients appearing in this expression are determined as  $\alpha_{4} = - 2R^2 f_{1}^{(2)}$,  $\alpha_{5} =R f_{3}^{(1)}$, $\beta_{2} = R^2 f_{2}^{(1)}$, and  $\gamma_2 = R^2 f_{4}^{(1)}$.   Plugging these solutions into the variations \eqref{eqn:gravitonVariations}, we find the following leading order variations of the metric perturbations
		\begin{subequations}\label{eqn:leadingGravitonVariations}
			\begin{align}
				\delta \partial_{u} H^{(0)}_{vv} ~ &= ~  - \dfrac{ 3\alpha_{1} u^{2}}{ 2R^{4}}  + 2 \dfrac{\alpha_{4}-\alpha_3}{R^2}~ , \label{eqn:HvvVarO1} \\
				\delta h_{v}^{+ (0)} ~ &= ~- \dfrac{3 \alpha_{1} u^{2}}{2 R^{2}} +  \frac{2\alpha_{2}u}{R} +\alpha_{3} + \beta_{2} ~ , \label{eqn:hplusvVarO1} \\
				\delta K^{(1)} ~ &= ~ \dfrac{\left(\ell^{2} + \ell + 1\right) \alpha_{1} u^{2}}{R^{4}} - \dfrac{\ell \left(\ell + 1\right) \alpha_{2} u}{R^{3}} + \dfrac{\alpha_{3} - \ell \left(\ell + 1 \right) \beta_{2}}{R^2} \, , \\
				\delta G^{(0)} ~ &= ~ - \dfrac{2\alpha_{1} u}{R^{2}} + \dfrac{2 \beta_{1}  }{R}~ , \label{eqn:GVarO1} \\
				\delta K^{(0)} ~ &= ~ \dfrac{\left(\ell^{2} + \ell + 1\right) \alpha_{1} u}{R^{2}} - \dfrac{\ell \left(\ell + 1\right) \beta_{1}}{R}  \label{eqn:KVarO1} \\
				\delta h_{v}^{- (0)} ~ &= ~ \gamma_2~ , \label{eqn:hminvVarO1} \\
				\delta h^{(0)}_{\Omega} ~ &= ~  2 R\gamma_1 ~ . \label{eqn:hOmegaVarO1} 
			\end{align}
		\end{subequations}
		It can be verified that these variations are consistent with the equations of motion we found in \eqref{eq:eomsol} and \eqref{eq:eomodd}.

		\subsection{Covariant charges}
		The gravitational charges associated with the asymptotic Killing vectors  \eqref{eqn:AKV} can be found using the covariant phase space formalism \cite{Iyer:1994ys, Barnich:2001jy} (also see \cite{Compere:2018aar} for a review):
		\begin{align}
			\cancel{\delta} Q_{\bar{\chi}}\left[\bar g_{\rho\sigma} ; h_{\rho\sigma}\right] ~ &= ~ \dfrac{\kappa}{16 \pi G} \int \left(\mathrm{d}^{2}x\right)_{\mu\nu} \sqrt{-\bar g} \left[ \bar{\chi}^{\nu} \nabla^{\mu} h - \bar{\chi}^{\mu} \nabla_{\sigma} h^{\mu\sigma} + \bar{\chi}_{\sigma} \nabla^{\nu} h^{\mu \sigma} + \dfrac{1}{2} h \nabla^{\nu} \bar{\chi}^{\mu} \right. \nonumber \\
			&\qquad\qquad\qquad\qquad\qquad\qquad \left. + \dfrac{1}{2} h^{\nu \sigma} \left(\nabla^{\mu} \bar{\chi}_{\sigma} - \nabla_{\sigma} \bar{\chi}^{\mu}\right) - \left(\mu \leftrightarrow \nu\right)  \right] \, ,
		\end{align}
		where $\bar{\chi}$ was defined in \eqref{eqn:genericDiffPWs}, $\cancel{\delta}$ indicates that the charges are not integrable in general and $\left(\mathrm{d}^{2}x\right)_{\mu\nu} = \frac{1}{4} \epsilon_{\mu\nu\rho\sigma} \mathrm{d}x^{\rho} \wedge \mathrm{d}x^{\sigma}$. Plugging in our solutions for the asymptotic Killing vectors from \secref{sec:AKV} and using the decomposition of the gravitational perturbations into partial waves as before, we find the charge in the even sector to be
		\begin{align}
			Q\left[\chi^{(0) \ell m}_{a}, \chi^{+ (0)}_{\ell m}\right] ~ &= ~ \dfrac{1}{2} \bigg[ \ell (\ell+1) \left(\partial_u {h_v^{+(0)}} \left( R \beta_1 -\alpha_1 u\right) 	+ 2 \alpha_1 {h_v^{+(0)}} \right) \nonumber \\
			&\qquad \qquad \qquad + 2 R^2 \alpha_{1} K^{(1)} + \left( 2 R \alpha_2 - 3 u \alpha_{1} \right) K^{(0)} \nonumber \\
			&\qquad \qquad \qquad + \left( u^2 \alpha_1 - 2 R u \alpha_2 - 2 R \alpha_2 	\right) \partial_u K^{(0)} \bigg] ~ .
		\end{align}
		Here, we used the following orthogonality relation for spherical harmonics to integrate over the two-sphere
		\begin{equation}\label{key}
			\int \dd \Omega D^AY(\Omega)_{\ell m}D_AY_{\ell' m'}(\Omega) ~ = ~ \ell(\ell+1)\delta_{\ell \ell'}\delta_{m,m'}.
		\end{equation}
		The $ \ell, m $ labels are again implicit in the charges.
		Upon imposing equations of motion \eqref{eq:eom}, this charge reduces to
		\begin{align}
			Q\left[\chi^{ \ell m}_{a}, \chi^{+ }_{\ell m}\right] ~ &= ~ \dfrac{1}{2}\bigg[ \ell \left(\ell + 1\right) \left(\partial_u {h_v^{+(0)}} \left(R \beta_{1} - 	\alpha_{1}  u\right) + 2 \alpha_{1} {h_v^{+(0)}} \right) \nonumber \\ 
			&\qquad \qquad \qquad - u \left( \ell \left(\ell + 1\right)  {c_h} + d_{K^{(0)}} \right) \alpha_{1} - \ell \left( \ell + 1\right) \alpha_{1} c_{K^{(0)}} \nonumber \\
			&\qquad \qquad \qquad + 2 R \left( \alpha _2 	d_{K_{(0)}} - R \alpha_3  c_{K^{(0)}} \right) \nonumber \\
			&\qquad \qquad \qquad + 2 \ell(\ell+1) \alpha_{1} d_h + 2 R^2 \alpha_{1} d_{K^{(1)}} \bigg] \, .
		\end{align}
		The first line of this expression contains free radiative data (which we chose to be $h^{+(0)}_{v}$), whereas the terms in the second line and below only contain non-radiative data. Such terms may be eliminated by an appropriate choice of boundary conditions that fix the integration constants appearing in the solutions to equations of motion in \eqref{eq:eomsol}.\footnote{Similar conditions were used in the study of higher dimensional supertranslations in \cite{Aggarwal:2018ilg}.} Similarly, in the odd sector, we find the following linearised charge
		\begin{align}
			Q\left[\chi^{- }_{\ell m}\right] ~ = ~ \dfrac{R}{2} \left[ \ell \left( \ell + 1\right) \gamma_1^{\ell, m} \partial_u h^{-(0)}_{v} \right] \, ,
		\end{align}
		where we used the following orthogonality relation to integrate over the two-sphere
		\begin{equation}\label{key}
			\int \dd \Omega ~ \epsilon^{AB}D_BY_{\ell m}(\Omega){\epsilon_{A}}^{C}D_CY_{\ell' m'}(\Omega) ~ = ~ \ell(\ell+1)\delta_{\ell \ell'}\delta_{m,m'}.
		\end{equation}
		We note that the Iyer-Wald \cite{Iyer:1994ys} and Barnich-Brandt \cite{Barnich:2001jy} charges coincide. Furthermore, there are no central extensions in the charge algebra; therefore the charge algebra is the same as the algebra of vector fields by the representation theorem of charge algebra \cite{Brown:1986ed,Compere:2018aar}. Moreover, since we are in the linearised theory, the algebra is abelian and thus the charge algebra is trivial at the linearised level.

		\subsection{More restrictive boundary conditions}\label{sec:restrictedBC}
		Some of the non-radiative data may be fixed by a slightly stronger boundary condition
		\begin{equation}\label{eqn:K0StrongerBC}
			\partial_u K^{(0)} ~ = ~ c_{K^{(0)}} ~ = ~ 0 \, .
		\end{equation}
		This condition implies that $ \alpha_1=0 $ in the asymptotic Killing vector as is evident from \eqref{eqn:KVarO1}. As we will see in \secref{sec:horizonAlgebra}, the most general asymptotic Killing vectors do not form a closed algebra under the usual Lie bracket if we go beyond the linearised theory. However, setting $ \alpha_1=0 $ indeed leads to a closed algebra under the usual Lie bracket even in the non-linear regime. Therefore, these boundary conditions are more natural in this sense.  The charge in the even sector then simplifies to
		\begin{align}
			Q\left[\chi^{ \ell m}_{a}, \chi^{+ }_{\ell m}\right] \Bigg|_{c_K^{(0)}=0,\alpha_1=0}~ &= ~ \dfrac{R}{2}\bigg[ \ell \left(\ell + 1\right) \left(\partial_u {h_v^{+(0)}} \beta_{1}   \right) + 2 \alpha _2 d_{K_{(0)}} 
			\bigg] \, .
		\end{align}
		\subsection{Physical interpretation of the charges}
		Out of the asymptotic Killing vectors, \eqref{eqn:AKVleadingOrder} and  \eqref{eqn:AKVsubleadingOrder}, some of them are exact isometries of the background.
		
		\begin{paragraph}{Entropy}
		One of them is the generator of Killing horizon 
		\begin{equation}
			\chi^S=\kappa \frac{\alpha_2^{\ell=0,m=0}}{R}(v\partial_v-u\partial_u)~.
		\end{equation}
		This is the generator of time translations and corresponds to scaling $u\rightarrow p u, v\rightarrow \frac{v}{p}$ for some constant $p$. The near-horizon charge corresponds to change in Bekenstein-Hawking entropy due to perturbation \eqref{eqn:gravitonModes}, $\delta S$, times the Hawking temperature, $T_H$, of the background Schwarzschild
		\begin{align}
			Q\left[\alpha_2^{\ell=0,m=0}=1\right]~ &= ~ {R} \delta R=\frac{\kappa}{2\pi} T_H \delta S
			\, .
		\end{align}
		Here, $\delta R:=d_{K_{(0)}}^{\ell=0,m=0} $, $T_H=\frac{1}{4\pi R}$, and $\delta S=\frac{8\pi^2 R^2 \delta R}{\kappa}$. This can be seen as follows. In our setup, the Schwarzschild metric together with the perturbation is \footnote{We only consider the perturbation with $d_{K_{(0)}}^{\ell=0,m=0}$ turned on for simplicity. The charge remains the same even for a more generic perturbation due to the choice of our vector field which is only sensitive to $d_{K_{(0)}}^{\ell=0,m=0}$.}
		\begin{equation}
			\dd s^2=A(u,v)\dd u\dd v+r^2\left(1+\kappa {\delta R}\right)\dd \Omega^2~.
		\end{equation}
	Therefore the area of the black hole changes from $4\pi R^2\rightarrow 4\pi R^2(1+\kappa\delta R)$ and the change in entropy due to the perturbation is $\frac{8\pi^2 R^2 \delta R}{\kappa}$. Recall that $\kappa=\sqrt{8\pi G}$.
\end{paragraph} 
\begin{paragraph}
	{Angular momentum}	
Another Killing vector correspond to the rotation symmetry of the Schwarzschild background, $\partial_\phi§$
\begin{equation} \label{eq:angmom}
{\chi}^L
~ = ~ \kappa R\left[ \gamma_1^{\ell=1, m=0} {\epsilon}^{A B} \partial_{B} Y_{\ell=1, m=0} \right]\partial_A=\frac{ \kappa R}{2}\sqrt{\frac 3 \pi} \gamma_1^{\ell=1, m=0}\partial_\phi\, .
\end{equation}
Here we used $Y_{\ell=1 m=0}=\frac{ 1}{2}\sqrt{\frac 3 \pi}\cos\theta$. The near-horizon charge corresponds to the angular momentum of the perturbation \eqref{eqn:gravitonModes} since the angular momentum of the background Schwarzschild is zero. The charge is
\begin{align} \label{eq:angmomCharge}
	Q\left[\gamma_1^{\ell=1,m=0}=\frac{ 1}{2}\sqrt{\frac 3 \pi}\right]~ &= ~ {R} \partial_u L(u)
	\, .
\end{align}
Here, $L(u):=\frac{ 1}{2}\sqrt{\frac 3 \pi}(h_v^{-(0)})^{\ell=1,m=0}$. This interpretation is motivated by the following argument. In our setup, the Schwarzschild metric together with the perturbation $L(u)$ turned on is \footnote{We only consider the perturbation with $L(u)$ turned on for simplicity. The charge remains the same even for a more generic perturbation due to the choice of our vector field which is only sensitive to $L(u)$.}
\begin{equation}
	\dd s^2=A(u,v)\dd u\dd v+r^2\dd \Omega^2-2\kappa L(u)\sin^2\theta ~ \dd v \dd \phi~.
\end{equation}
If $L(u)=u a$ for some constant $a$, this is the metric of Kerr black hole with a small rotation parameter $\kappa a$ that has been linearised around Schwarzschild black hole by keeping only the terms linear in $\kappa a$. Then the charge \eqref{eq:angmomCharge}, $Q=R a$, is easily seen to be related to the angular momentum of such a black hole.
\end{paragraph}

		\section{Symmetry algebra} \label{sec:algebra}
		In this section, we study the algebra of the near-horizon symmetries we found in \secref{sec:AKV} and its closed sub-algebras, and the algebra of the symmetries (in \eqref{eqn:KV}) to all orders in $ v $. We examine the closure of these algebras under the usual Lie bracket and determine necessary conditions for closure beyond the linearised regime. Of the two resulting closed sub-algebras of the near-horizon asymptotic Killing symmetries, one corresponds to a maximal sub-algebra while the other is the dominant one for large black holes. It must be noted that in the linearised theory, the algebra is abelian and therefore always closes. This can be seen from \eqref{eqn:genericDiffPWs} by noting that $ \bar \chi $ is $ \mathcal O (\kappa) $. Therefore, the Lie bracket is $ \mathcal O(\kappa^2) $. So, all the statements we make below about the closure (or otherwise) go beyond the linear approximation and are only relevant when one works with the full non-linear theory. In the non-linear theory, we expect that Barnich-Troessaert bracket \cite{Barnich:2011mi} would lead to closure of the full algebra.

		\subsection{Near-horizon symmetry algebra}\label{sec:horizonAlgebra}
		From the solutions for the asymptotic Killing vectors in \eqref{eqn:AKVleadingOrder} written in terms of arbitrary partial wave dependent constants, we notice that the complete asymptotic Killing vector can be parameterised in terms of arbitrary functions $F\left(\Omega\right)$, $M\left(\Omega\right)$, $P\left(\Omega\right)$ and $Z^{A} \left(\Omega\right)$ on the unit sphere as follows:
		\begin{align}\label{eqn:AKV}
	{\bar{\chi}}\left[ F, M, P, Z^{A}\right] ~ &= ~\kappa\bigg[ \left( F \left(\Omega\right) \dfrac{u^2}{2 R^{2}} + M \left(\Omega\right) \dfrac{u}{R} + P \left(\Omega\right) \right) \partial_u - F \left(\Omega\right) \partial_v - \frac{u}{R^2}\gamma^{AC} \partial_{C} F \left(\Omega\right)  \partial_{A} \nonumber \\
			&\qquad + \frac{Z^A \left(\Omega\right)}{R} \partial_A + ~ \cdots \bigg]~ ,
		\end{align}
		with the dots representing sub-leading terms and $ \Omega $ a point on the unit $2$-sphere.
		Here,
		\begin{eqnarray}\label{eq:AKVsphere}
			F(\Omega)&=& 
			\sum\limits_{\ell,m}\alpha_1^{\ell m}Y_{\ell m}(\Omega)~,\cr
			M(\Omega)&=&-\sum\limits_{\ell,m}\alpha_2^{\ell m}Y_{\ell m}(\Omega)~,\cr
			P(\Omega)&=&-\sum\limits_{\ell,m}\alpha_3^{\ell m}Y_{\ell m}(\Omega)~,\cr
			Z^A(\Omega)&=&\sum\limits_{\ell,m}\beta_1^{\ell m}\gamma^{AC}D_CY_{\ell m}(\Omega)+\gamma_1^{\ell m}\epsilon^{AB}D_BY_{\ell m}(\Omega)~.
		\end{eqnarray}
		
		Note that the spacetime index of the asymptotic  Killing vector \eqref{eqn:AKVleadingOrder}  must be raised with the background metric ($\bar{g}^{\mu\nu} \bar{\chi}_{\nu}$) to arrive at this expression. 
		The even and odd modes on the sphere result in independent even and odd functions on the sphere which have been combined into two arbitrary functions $Z^{A}\left(\Omega\right)$ on the sphere.  It can be checked that the lie bracket of \eqref{eqn:AKV} 
		does not close,
			\begin{align}\label{eq:algebra}
			[\bar\chi(F_1,M_1,P_1,Z^A_1),~\bar \chi(F_2,M_2,P_2,Z^A_2)] ~ &\neq ~ \bar{\chi} \left(F_{12},M_{12},P_{12},Z^A_{12}\right)~,
		\end{align}
		for any arbitrary functions $ F_{12},~ M_{12},~P_{12},~Z_{12} $. However, this is a non-linear statement since the right hand side is $ \mathcal O(\kappa^2) $.
		 
		It is conceivable that the non-linear closure of this algebra requires the addition of more terms to the asymptotic Killing vector \eqref{eqn:AKV} that may resemble \cite{Adami:2021nnf,Chandrasekaran:2023vzb, Odak:2023pga} where diffeomorphisms along the horizon (in addition to $ {\rm Diff }(S^2) $) were found. Indeed, it can be checked that the Lie bracket \eqref{eq:algebra} generates terms of the type $ { F}_{n}(\Omega)u^n\partial_u$ for some arbitrary function on the sphere $ F_n(\Omega) $ and an arbitrary integer $ n $. We defer further discussion to \secref{sec:conclusions}.

		\subsubsection{Maximal closed sub-algebra}
		The maximal sub-algebra that closes is that of vector fields with $F\left(\Omega\right) = 0$.
		
		\begin{align}\label{eq:algebraFzero}
			[\bar\chi(M_1,P_1,Z^A_1),~\bar \chi(M_2,P_2,Z^A_2)]  = \bar{\chi} \left(M_{12},P_{12},Z^A_{12}\right)
		\end{align}
		with
		\begin{subequations}
			\begin{align}
				M_{12} ~ &= ~ \kappa\left[\dfrac{Z^{A}_{1} \partial_{A} M_{2} - Z^{A}_{2} \partial_{A} M_{1}}{R}\right]~, \\
				P_{12} ~ &= ~\kappa\bigg[ M_{1} P_{2} - M_{2} P_{1} +\frac{ Z^{A}_{1} \partial_{A} P_{2} - Z^{A}_{2} \partial_{A} P_{1}}{R}\bigg ]~, \\
				Z^{A}_{12} ~ &= ~\kappa\left [ \frac{Z^{B}_{1} \partial_{B} Z^{A}_{2} - Z^{B}_{2} \partial_{B} Z^{A}_{1}}{R} \right ]\, .
			\end{align}	
		\end{subequations}
		As pointed out in \secref{sec:restrictedBC}, this sub-algebra  naturally arises from a stronger boundary condition \eqref{eqn:K0StrongerBC} than finiteness of metric perturbations on the horizon. It can be verified that it consists of diffeomorphisms of the two-sphere, $\text{Diff}\left(S^{2}\right)$ generated by $Z^{A}\left(\Omega\right)$, in semi-direct sum with two supertranslations, generated by $M\left(\Omega\right)$ and $P\left(\Omega\right)$ on the horizon. This is in agreement with the results of \cite{Chandrasekaran:2018aop} and also the results found in \cite{Donnay:2016ejv} where two copies of the Virasoro algebra were found and the complete set of $\text{Diff}\left(S^{2}\right)$ was anticipated. 
		
		It is interesting to note that in our setup, in contrast to the analysis of \cite{Chandrasekaran:2018aop,Donnay:2016ejv}, we see that one of the supertranslations (associated with $M\left(\Omega\right)$) and $ {\rm Diff} (S^2) $ are suppressed in powers of $ R $ for a large black hole, $ R\gg \kappa $.

		\subsubsection{Symmetries of large black holes}
		While we saw that $ F(\Omega)=0 $ was necessary for the maximal closed algebra defined by the conventional Lie bracket, it is evident that for very large black holes, the Killing vector \eqref{eqn:AKV} reduces to
		\begin{align}\label{eqn:AKVlargeR}
			\bar{\chi} \left[ F, P\right] ~ &= ~\kappa\bigg[ P \left(\Omega\right) \partial_{u} - F\left(\Omega\right) \partial_{v} + \mathcal{O}\left(\dfrac{1}{R}\right)\bigg] \, .
		\end{align}
		The algebra generated by this asymptotic Killing vector clearly closes with a trivial commutator leading to an algebra of two copies of $ u(1) $. In the maximal sub-algebra discussed in the previous subsection, the near-horizon symmetry associated with $F\left(\Omega\right)$ did not contribute. However, we see that for a large black hole this symmetry dominates and without any obstructions to the algebra closure. Thus, it  has physical significance for scattering processes near large black holes which are of semi-classical interest.

		\subsection{Symmetry algebra to all orders in $ v $}\label{sec:globalAlgebra}
		Our boundary conditions allow us to write the asymptotic Killing vector fields  \eqref{eqn:AKV} to all order in $ v $ using \eqref{eqn:KV}:
		\begin{align}
			{\bar{\chi}} ~ &= ~  \kappa \bigg[ \left(- \dfrac{\hat{\bf F} \left(v, \Omega\right)}{A\left(r\right)} + \dfrac{\hat{A}(u,v)}{A\left(r\right)} \partial_{v} {\bf F} \left(v, \Omega\right) +{\bf  F} \left(v, \Omega\right) \dfrac{\partial_{v} \hat{A}(u,v)}{A\left(r\right)} \right) \partial_u -{ \bf F} \left(v, \Omega\right) \partial_v\nonumber \\ 
			&\qquad \qquad \qquad - \tilde{A}(u,v)~ \gamma^{AC}{\partial_{C} {\bf F} \left(v, \Omega\right)}\partial_{A} 
			+ {{\bf Z}^A\left(v, \Omega\right) } \partial_A \bigg]~ \,
		\end{align}
		where we defined the background functions
		\begin{align}
			\hat{A}(u,v) ~ \coloneqq ~  \int \mathrm{d}u \, A\left(r\right) \, , \quad \text{and} \quad \tilde{A} (u,v)~ \coloneqq ~  \int \mathrm{d}u \, \dfrac{A\left(r\right)}{r^{2}} \, .
		\end{align}
		Here,
		\begin{eqnarray}\label{eq:AKVspherefull}
			{	\bf F}(v,\Omega)&=& 
			\sum\limits_{\ell,m}f_1(v)^{\ell m}Y_{\ell m}(\Omega)~,\cr
			\hat {\bf F}(v,\Omega)&=&\sum\limits_{\ell,m}f_3(v)^{\ell m}Y_{\ell m}(\Omega)~,\cr
			{\bf Z}^A(v,\Omega)&=&\sum\limits_{\ell,m}f_2(v)^{\ell m}\gamma^{AC}D_CY_{\ell m}(\Omega)+f_4(v)^{\ell m}\epsilon^{AB}D_BY_{\ell m}(\Omega)~.
		\end{eqnarray}
		
		The algebra of these vector fields does not close in general (since the asymptotic algebra did not close \eqref{eq:algebra})
		\begin{align}
			[\bar\chi(\hat{\bf F}_1, {\bf F}_1, ,{\bf Z}^A_1),~\bar \chi(\hat{\bf F}_2, {\bf F}_2, {\bf Z}^A_2)] ~ &\neq ~ \bar{\chi} \left(\hat{\bf F}_{12}, {\bf F}_{12}, {\bf Z}^A_{12}\right)~.
		\end{align}
		However, it can easily be checked that the sub-algebra formed by vector fields with ${\bf F}\left(v, \Omega\right) = 0$ does indeed close with
		\begin{align}
			\hat{\bf F}_{12} ~ = ~ \kappa\left [{{\bf Z}^{A}_{1} \partial_{A} \hat{\bf F}_{2} - {\bf Z}^{A}_{2} \partial_{A} \hat{\bf F}_{1}}\right ] \quad \text{and} \quad{\bf  Z}^{A}_{12} ~ = ~\kappa\left [ {\bf Z}^{B}_{1} \partial_{B} {\bf Z}^{A}_{2} -{\bf  Z}^{B}_{2} \partial_{B} {\bf Z}^{A}_{1} \right ]\, .
		\end{align}
		Note that this sub-algebra coincides with the one formed by \eqref{eqn:AKV} for $ F=M=0 $, in the near-horizon region. Therefore, it consists of only one supertranslation instead of two.  Notice that the Killing vector now contains arbitrary functions of $v$ and a dependence on $u$ via the background function $A\left(r\right)$.  It would be interesting to find less stringent restrictions on the function ${\bf F}\left(v, \Omega\right)$ that may still lead to a closed algebra. In general, it is also of interest to find the general role and meaning of these symmetries that do not lead to a closed algebra, both on the horizon to leading order and also possibly to all orders in the distance to the horizon, parametrised by $v$.

		\section{Conclusions and outlook} \label{sec:conclusions}
		In this article, we studied the perturbations of the Schwarzschild black hole in a partial wave basis in a gauge that allows for radiation crossing the horizon. This setup is ideally suited for studying near-horizon scattering. We found the residual symmetries that preserve the aforementioned ``radiation gauge''.  By requiring only that the perturbations be finite on the horizon, we found the Killing vectors not only near the horizon but also to all orders in $v$ in \secref{sec:globalAlgebra}. Remarkably, with no further restrictions we found the charges on the horizon corresponding to these symmetries to be finite and non-vanishing. The physical significance of the sub-leading terms is not clear as they do not contribute to the charges. Nevertheless, there is some evidence that such sub-leading terms can be related to sub-leading soft theorems for scattering in flat spacetime \cite{Conde:2016csj, Conde:2016rom}. Our leading order vector fields correspond to near-horizon symmetries that were not previously known. In the linearised theory, the algebra is abelian and therefore closes to leading order. However, it is curious that the complete algebra on the horizon does not close with the conventional Lie bracket owing to sub-leading non-linear obstructions. We identified two important sub-cases that lead to a closed algebra. The first is obtained by setting $ F(\Omega)=0 $ and leads to an algebra  of two supertranslations in semi-direct sum with all diffeomorphisms of the two sphere, $\text{Diff}\left(S^{2}\right)$. We found a specific restriction on the boundary conditions that naturally leads to this maximal sub-algebra. The second sub-algebra emerges in the large black hole limit and is considerably smaller containing only two copies of $ u(1) $. It would be interesting to check if a modified bracket, like the one due to Barnich and Troessaert \cite{Barnich:2011mi, Barnich:2013sxa}, closes the complete algebra with no restrictions.\footnote{AA would like to thank Glenn Barnich for insightful discussions related to the modified bracket.}  
		
		\paragraph{Comparison to previous work: } Several authors have studied the symmetries of the black hole horizon in recent years as mentioned in the introduction. Here we list a comparison between some of these works and our results. In \cite{Donnay:2016ejv}, the gauge that was used is different from the radiation gauge that we employ. Consequently, we find different near-horizon symmetries. In \cite{Hawking:2016sgy}, the choice of gauge, fixed the size of the horizon. In our paper, we found that some symmetries do change the size of the black hole, while others affect its shape. In \cite{Adami:2020amw, Chandrasekaran:2018aop, Odak:2023pga}, further restrictions on the symmetries were imposed in order to, for instance, preserve the location of the (bifurcate) horizon. In this work, we allow for symmetries that may change the location of the horizon. Indeed, scattering processes near the black hole horizon may change the location of the horizon in general. In the linearised theory, the perturbations (including but not limited to changes in the location of the horizon) are small when $\kappa/R \ll 1$. However, we may improve our results order by order in $\kappa$, thereby generating non-linear effects, all the while preserving our relaxed boundary conditions. It would be interesting to compare the results of such non-linear effects with the recent work of \cite{Chandrasekaran:2023vzb, Odak:2023pga}. In particular, it is of interest to compare  the resulting (potentially) closed algebra (by adding more terms to our vector fields) with what is found there. The maximal sub-algebra that does close is in line with the previous literature \cite{Donnay:2016ejv, Chandrasekaran:2018aop}. However, we found an additional symmetry which finds its place in a smaller commuting sub-algebra that survives in the large black hole limit. Finally, in contrast to all the above works, we find a hierarchy between the different supertranslations in the large black hole limit.
		
		The primary motivation for studying linear perturbations in the partial wave basis, as opposed to the non-linear theory, is that scattering processes in the near-horizon region arise naturally in this basis \cite{Gaddam:2020rxb, Gaddam:2020mwe, Betzios:2020xuj, Gaddam:2021zka, Gaddam:2022pnb}. The results of this article can therefore to be seen as capturing all the symmetries of near-horizon scattering. The next natural step is to study the  implications of these symmetries for near-horizon scattering.  This warrants an understanding of the Ward-Takahashi identities of these near-horizon symmetries. The relation between these identities and emergent soft graviton theorems near the horizon is an important question that we will report on in an upcoming work \cite{us}. 
		
		In a dynamical problem of collapse, boundary conditions imposed on past null infinity would in-principle dictate boundary conditions to be imposed on the horizon. Similarly, if the black hole is to evaporate in its entirety, boundary conditions on future null infinity would also determine those on the horizon. However, determining these boundary conditions on the horizon in practice is a difficult task. One may approach this by setting up a WKB-like analysis where equations of motion are solved order by order both near the horizon and near infinity. An appropriate matching condition in the intermediate region will then determine boundary conditions on the horizon in terms of the choices at infinity.\footnote{NG would like to thank Godwin Martin for discussions regarding this idea.} This might be possible in our formalism because the symmetries are known to all orders in $ v $ and therefore at any point in spacetime.
		
		Recently, in \cite{He:2023qha}, a scattering algebra of 't Hooft's associated with shockwaves was shown to be related to the soft algebra near infinity. This can be exploited to provide a physical interpretation of the familiar antipodal matching condition at spatial infinity from the perspective of scattering in the bulk \cite{GaddamHe}. Our results in this paper, especially given their direct relevance for near-horizon scattering and an analogous shockwave algebra in black hole backgrounds, may potentially imply a similar antipodal matching condition on the bifurcation sphere.\footnote{A primitive, and possibly inaccurate, version of such an antipodal identification was anticipated and discussed in \cite{Hooft:2016itl, Betzios:2017krj, Betzios:2020wcv, tHooft:2022bgo}.}
		
		While we have entirely focused on gravitational perturbations in this paper, it is also possible to study charged particle scattering near the black hole horizon \cite{FGG}. Our analysis can be extended in a straightforward manner\footnote{See also \cite{Gerlach:1980tx}.} to study emergent and potentially new symmetries of QED near the black hole horizon and ask about their relationship with an emergent soft-photon theorem near the horizon.

		Asymptotic symmetries near null infinity are subtle in different dimensions {\cite{Hollands:2016oma,Hollands:2004ac,Kapec:2015vwa, Pate:2017fgt, Aggarwal:2018ilg, Colferai:2020rte, Campoleoni:2020ejn, Chandrasekaran:2021vyu, Chowdhury:2022gib, Chowdhury:2022nus, Capone:2021aas, Capone:2021ouo, Fuentealba:2021yvo,Fuentealba:2022yqt,Lionetti:2022lwu,Capone:2023roc} because the desirable asymptotic charges seem to diverge in a naive $ r\rightarrow\infty $ limit. Such divergences do not arise in the case of higher dimensional black holes where the near-horizon charges are computed in the $ r\rightarrow R $ limit. Therefore, our formalism is easily adapted to the case of higher dimensional black holes.\footnote{A similar observation was made in \cite{Akhmedov:2017ftb}.} An appropriate choice of spherical harmonics would imply that in the final expressions, only the eigenvalues would have to be adapted in comparison to our results. Further potential subtleties, if any, may be interesting to understand. 
			
			Finally, it is conceivable that a version of the memory effect in the near horizon region owing to the symmetries near the horizon may have observable consequences. For instance, it would be of great importance to understand whether the radiation emerging from the horizon can leave an imprint on the luminosity fluctuations of the spectra of stellar oscillations (of the S-stars orbiting the Saggittarius A* for example). While there are other (potentially stronger) sources for the said luminosity fluctuations like those due to internal pressure and density dynamics of the orbiting stars, those owed to near-horizon radiation considered in this paper are likely to occur on a much shorter time-scale (inverse of the speed-of-light as opposed to the inverse of the speed-of-sound). This separation of scales may make these signals potentially observable in the near future. However, such measurements depend greatly on an asteroseismological understanding of the spectral type, or any other signature, that captures the average and/or dominant luminosity fluctuations of the stellar oscillations of interest.

			\section*{Acknowledgements}
			It is a pleasure to thank Anupam A.H., Uddipan Banik, Glenn Barnich, Chandramouli Chowdhury, Fabiano Feleppa, Nico Groenenboom, and Alok Laddha for various helpful discussions. We are also grateful to the Cargese Summer School 2021 where this collaboration bore fruit, and Perimeter Institute, where a part of this work was carried out, for hosting both authors. 
			
			AA is a Research Fellow of the Fonds de la Recherche Scientifique F.R.S.-FNRS (Belgium). AA is partially supported by IISN – Belgium (convention 4.4503.15) and by the Delta ITP consortium, a program of the NWO that is funded by the Dutch Ministry of Education, Culture and Science (OCW).
			
			NG was supported by the Delta-Institute for Theoretical Physics (D-ITP) that is funded by the Dutch Ministry of Education, Culture and Science (OCW) during the initial stages of this work and is currently supported by project RTI4001 of the Department of Atomic Energy, Govt. of India.

			\begin{appendix}

			\end{appendix}
			
			\bibliographystyle{ytphys}
			\bibliography{references}
			
		\end{document}